\documentstyle[psfig]{elsart}
\begin{document}

\begin{center}
{\Large Reducibility, Arrhenius plots and the Uroboros 
Dragon\footnote{
The
Uroboros dragon, or the snake eating its own tail, is an archetype that can be
traced from the Egyptians through the Greeks and the Gnostic religions into the
Alchemical writings of the Middle Ages and Renaissance all the way to modern
times. A symbol of self reference (think of Escher's hand drawing the hand
that draws it...), it has been used to explain the universe. It has been
applied by the authors of the preprint with a somewhat more modest intent.},\\
a reply to the preprint \\
``Correlations in Nuclear Arrhenius-Type Plots'' \\
by M.B. Tsang and P. Danielewicz.\\}

\vspace{1.0cm}
{\large L.G. Moretto, L. Beaulieu, L. Phair, and G.J. Wozniak\\}

\vspace{0.6cm}
{\large\em Nuclear Science Division, Lawrence Berkeley National Laboratory,
Berkeley, CA 94720, USA\\}

\vspace{0.6cm}
\date{\today}
June 5, 1997

\end{center}

\vspace{0.5cm}

In previous publications \cite{Moretto95,Tso95,Moretto97}, 
which have generated more attention than
anticipated \cite{Bot95,Del95,Gro97,Ara97,Tok97,Tsa97}, 
we showed that the
$n$-fragment multifragmentation probabilities
at any given transverse energy $P_n(E_t)$, could be reduced by means of the
binomial expression 
\begin{equation}
P_n^m=\frac{m!}{n!(m-n)!}p^n\left(1-p\right)^{m-n}
\end{equation}
to a single one fragment probability $p(E_t)$, where $E_t=\sum E_i\sin ^2\theta
_i$, the sum of the kinetic energies of all charged particles in an event
weighted by the sine squared of their polar angle. This
``\underline{reducibility}'' with $E_t$ 
was taken as a strong indication that fragments were emitted nearly
independently of one another with the same probability  $p(E_t)$. We further
noticed that, under the assumption that $E_t$ is proportional to the excitation
energy $E$ (and thus temperature 
$T\propto\sqrt{E_t}$), the Arrhenius plots of $\log(1/p)$
versus $1/\sqrt{E_t}$, produced straight lines. We called this 
``\underline{thermal
scaling}'' because it suggests the Boltzmann form for the elementary probability
$p\propto\exp (-B/T)$ or
\begin{equation}
\log\frac{1}{p}=\frac{B}{T}=\frac{B'}{\sqrt{E_t}},
\end{equation}
with $B$ representing the barrier for fragment emission.

The present 
preprint
attacks both of our claims. The first on the basis 
 that
our binomial fits could be matched in quality by a constrained Poissonian
function. The second on the basis that the Arrhenius plots are allegedly shown 
to be the result of
a strong autocorrelation.

Since the second criticism is potentially more devastating than the first, 
we begin with it.

As a preemptive strike, 
the authors raise concerns about the relevance of using $\sqrt{E_t}$
as a measure of the temperature. They cite as evidence to the contrary recent
work exploring the impact parameter dependence of temperatures 
extracted from excited state
populations \cite{Hua97}. In that work a value of $T$=4.6 MeV is determined,
largely independent of impact parameter for moderately central to very central
collisions of Au+Au at 35 MeV/nucleon. 
We are not unduly concerned by this result since temperatures of
4-5 MeV are almost {\em always} extracted from such studies, independent of
bombarding energy \cite{Che87,Mor94}, entrance channel mass asymmetry 
\cite{Che87,Mor94},
as well as impact parameter \cite{Hua97,Zhu92}. Since such extracted 
temperatures seem to be insensitive to {\em most} 
features of heavy-ion reactions,
we would not be surprised to find that they are also insensitive to $E_t$.

Furthermore, temperatures 
extracted from the double isotope ratio method \cite{Alb85} do indeed
show a broad range of temperatures for heavy ion reactions at
intermediate energies \cite{Ma97}.

Coming to the heart of their criticism, the authors argue that the observed
linearity of the Arrhenius plots is due to autocorrelation. However,  
a fatal flaw mars the derivation of the proposed
explanation for linear Arrhenius plots \cite{Tsa97} and undermines the
conclusions of the preprint. This flaw is 
\protect\underline{circular logic in the
algebra}, leading to a trivial identity that always holds, regardless of whether
or not there exists an autocorrelation.

Assuming with the authors that the binomial number of ``throws''
$m$ is constant (for many experiments this is
 a good assumption) 
we consider
the Arrhenius plot constructed with the average intermediate mass fragment
(IMF) multiplicity $\left<
n\right>=mp$ and $E_t$. So we plot $\log 1/\left< n\right>$ 
versus $1/\sqrt{E_t}$ and in all
experimental cases we
obtain a straight line. As we have observed \cite{Moretto97}, 
the authors also observe that $E_t$
is constructed \underline{both} with the light charged particles and the IMFs.
In the authors' notation
\begin{equation}
E_t=\left(N_C-\left< n\right>\right)E_t^{\rm LP}+\left< n\right>E_t^{\rm IMF},
\end{equation}
where $N_C$ is the number of charged particles in an event and $E_t^{\rm
LP}$ and $E_t^{\rm IMF}$ are the average transverse energies of the light
particles and IMFs, respectively.                                  
The use of $\left< n\right>$ to construct this independent variable $E_t$
creates the danger of autocorrelation in the study of the dependence of
$\left< n\right>$ on $E_t$ (as in the Arrhenius plot). This we also have 
discussed
extensively \cite{Moretto97}.

Now the authors \cite{Tsa97} observe that $\left< n\right>$ and $N_C$ are
\underline{globally} very strongly correlated. They fit this correlation with
the linear form
\begin{equation}
N_C=a'+b'\left< n\right>
\label{eq:Nc_Nimf}
\end{equation}
and finally they express $E_t$ as a function of the single variable
$\left< n\right>$
\begin{equation}
E_t=b\left(\frac{a}{b}+\left< n\right>\right)
\label{eq:auto_corr}
\end{equation}
where $a=a'E_t^{\rm LP}$ and $b=(b'-1)E_t^{\rm LP}+E_t^{\rm IMF}$. 


\begin{figure}
\centerline{\psfig{file=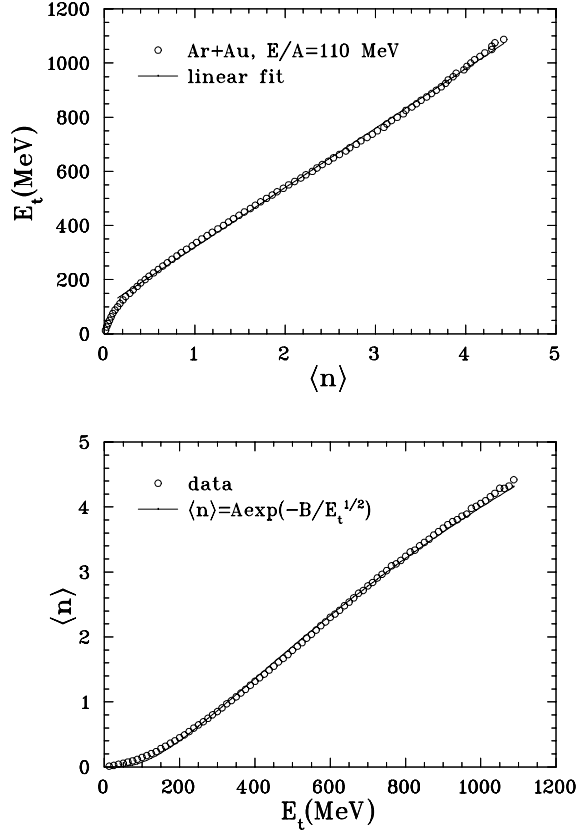,angle=90,height=11.0cm}}
\caption{Top panel: The transverse energy $E_t$ as a function of the average
IMF multiplicity $\left< n\right>$ for the reaction Ar+Au, E/A=110 MeV. The
line is a straight-line fit, Eq.~(\protect\ref{eq:auto_corr}). Bottom panel:
$\left< n\right>$ as a function of $E_t$. The fit is the exponential fit of
Eq.~(\protect\ref{eq:fprime}) used in refs.
\protect\cite{Moretto95,Tso95,Moretto97}.}
\label{fig:Nimf_Et}
\end{figure}

A quicker but nearly 
equivalent procedure would have been to fit the equally strongly
correlated quantities $E_t$, $\left< n\right>$ directly with the above form
(see the top panel of Fig.~\ref{fig:Nimf_Et})
since $N_C$ and $E_t$ are very tightly correlated as well. The values of $a$
and $b$ from a direct fit using Eq.~(\ref{eq:auto_corr}) 
are 100 MeV and 220 MeV respectively. These values are
nearly equal to those extracted by the authors ($a$=106 MeV, $b$=213 MeV)
when starting with Eq.~(\ref{eq:Nc_Nimf}).

A moment's
reflection shows that Eq.~(\ref{eq:auto_corr}) contains the original
experimental dependence of $\left< n\right>$ on $E_t$ (bottom panel of
Fig.~\ref{fig:Nimf_Et}), and it does not differ
in any significant way, other than in the linear fit, from the data used to
construct the Arrhenius plots. \underline{It is just its inverse!} Here is the
start of the \protect\underline{vicious circle}.

In fact, to fit the data they could have used with better results
the Boltzmann-like functional form which gives linear Arrhenius plots
as we have done in the bottom panel of Fig.~\ref{fig:Nimf_Et}. One can argue on
which of the two is a better fit, but not on the fact that, 
for the two functions,
one is approximately the inverse of the other.

Armed with Eq.~(\ref{eq:auto_corr}), the authors take an experimental Arrhenius
plot which is linear, and they substitute $E_t$ with the value calculated from
Eq.~(\ref{eq:auto_corr}). \protect\underline{Since the linear form is 
a good approximation to
$E_t$}, nothing much happens to \protect\underline{the Arrhenius plot} 
which of course 
\protect\underline{remains linear}.
It simply means that the
form of Eq.~(\ref{eq:auto_corr}) is a good approximation to the data!

Yet the authors make the surprising claim 
that the linearity in such a plot is the result of an
autocorrelation because now one has $\left< n\right>$ both in the ordinate (as 
$\log 1/\left< n\right>$) and in the abscissa
(as $1/\sqrt{b(a/b+\left< n\right>)}$).

In what follows 
we show that this technique for demonstrating autocorrelations is
so 
general and fundamentally flawed, 
that it will give the {\em same} result for all 
functions, {\em even for those in which autocorrelation has been ruled out by
construction.} We call this method the Uroboros method, for reasons that will
become clear below.

As we noticed above, Eq.~(\ref{eq:auto_corr}), either fitted directly to the
data or indirectly through Eq.~(\ref{eq:Nc_Nimf}), gives an empirical
approximation to the \underline{total} dependence of $E_t$ upon 
$\left< n\right>$
\begin{equation}
E_t=g\left(\left< n\right>\right).
\label{eq:total_dep}
\end{equation}
On the other hand, we have the \protect\underline{total} dependence of $\left< n\right>$ upon $E_t$
directly from the data
\begin{equation}
\left< n\right>=f\left(E_t\right).
\label{eq:ave_n}
\end{equation}
Clearly $g\left(\left< n\right>\right)$ is an
approximation to the inverse of $f\left(E_t\right)$, $g\approx f^{-1}$. Now we
construct the Arrhenius plot
\begin{equation}
F\left(\left< n\right>\right)=F\left(f'\left(E_t\right)\right)
\end{equation}
with 
\begin{equation}
\left< n\right>=f'\left(E_t\right)\propto\exp\left(\frac{-K}{\sqrt{E_t}}\right).
\label{eq:fprime}
\end{equation}

If the Arrhenius plot is linear it means the $f'(E_t)\equiv f$. Hence,
\begin{equation}
F\left(\left< n\right>\right)=F\left(f\left(E_t\right)\right)=
F\left(f\left(g\left(\left< n\right>\right)\right)\right)=
F\left(f\left(f^{-1}\left(\left< n\right>\right)\right)\right)=
F\left(\left< n\right>\right)
\end{equation}
which is an identity, in the limit of $g=f^{-1}$.

The \protect\underline{vicious circle of circular logic}
can be summarized as follows: given any function $y=y(x)$,
substitute in it $x$ with its value $x(y)$: $y=y(x(y))$. This gives always the
identity $y=y$.

This result proves, if there was any doubt, 
that \underline{any} function is completely correlated with its own
inverse, and it is a great source of straight lines, irrespective of whether
there is an autocorrelation or not.

The flaw, in our specific case, lies with 
the fact that Eq.~(\ref{eq:total_dep}) gives the
{\em total empirical} 
dependence of $E_t$ on $\left< n\right>$ {\em irrespective of
whether there is autocorrelation or not} and consequently {\em it is the
approximate inverse of Eq.~(\ref{eq:ave_n}) and of Eq.~(\ref{eq:fprime})}.
Hence the Uroboros, the snake eating its tail.

It is left for the diligent reader to test the universality of this method by
applying the Uroboros to discover autocorrelations in 
the most venerable dependences in physics.

It is clear from the previous considerations that the authors' vicious circle 
does not
prove anything, least of all autocorrelation.

In order to appreciate the true role of autocorrelations, let as before
\begin{equation}
\left< n\right>=f\left(E_t\right)
\end{equation}
and 
\begin{equation}
E_t=E_t\left(E,\left< n\right>\right)
\end{equation}
where $E$ is the excitation energy of the system. The dependence of 
$\left< n\right>$ on $E_t$ can be best explored differentially
\begin{equation}
\Delta\left< n\right> =\frac{d\left< n\right>}{dE_t}\Delta E_t.
\label{eq:n_dep}
\end{equation}
Now, the change in transverse energy can be written
\begin{equation}
\Delta E_t=\left.\frac{\partial E_t}{\partial E}\right|_{n} 
\Delta E + \left.\frac{\partial E_t}{\partial\left< n\right>}\right|_{E}
\Delta\left< n\right>
\label{eq:differential}
\end{equation}
where the first term represents the 
change in $E_t$ due to the change in excitation
energy 
and the second term represents the trivial change
in $E_t$ due to autocorrelation. 
The \underline{partial} derivative 
$\left.\frac{\partial E_t}{\partial\left< n\right>}\right|_E$ at
\underline{constant} E (not the \underline{total} derivative) tells us about the
autocorrelation. 

Substituting Eq.~(\ref{eq:differential}) in 
Eq.~(\ref{eq:n_dep}) we have 
\begin{equation}
\Delta\left< n\right>\left(1-\frac{d\left< n\right>}{dE_t}
\left.\frac{\partial E_t}{\partial\left< n\right>}\right|_E\right)=
\frac{d\left< n\right>}{dE_t}
\left.\frac{\partial E_t}{\partial E}\right|_{\left< n\right>}\Delta
E
\end{equation}
where the complete left hand side is the increase in $\left< n\right>$ induced
by $\Delta E$ 
\protect\underline{corrected for the autocorrelation}. Failure to appreciate the
difference between $\left.\frac{\partial E_t}{\partial\left< n\right>}\right|_E$
and $\frac{dE_t}{d\left< n\right>}$ leads to disastrous 
\protect\underline{circular reasoning}.

In fact, if we write
\begin{equation}
\Delta E_t=\frac{dE_t}{d\left< n\right>}\Delta\left< n\right>
\label{eq:total_derivative}
\end{equation}
and substitute in Eq.~(\ref{eq:n_dep}) we obtain as the authors did
\begin{equation}
\Delta\left< n\right>=\frac{d\left< n\right>}{dE_t}
\frac{dE_t}{d\left< n\right>}\Delta\left< n\right>=\Delta\left< n\right>,
\end{equation}
or the Uroboros (again).

Of course, the authors might argue that they know the first term of
Eq.~(\ref{eq:differential}) to be close to zero because they believe the
temperature changes at most by 1 MeV. Then indeed only the second term, or the
autocorrelation would survive.
This position is clearly untenable. In fact, were the authors correct, the
derivative in Eq.~(\ref{eq:total_derivative}) (which would now be the same as
that in the second term of Eq.~(\ref{eq:differential})) should have the value
of approximately 20-40 MeV per IMF, which is the measured 
mean transverse energy per
IMF \cite{Des93,Phair97} for this particular system. Instead, it has
the enormous value of 220 MeV per IMF! (See Fig.~\ref{fig:Nimf_Et}.) This
alone should make the authors wonder and worry about constant temperature.
In any case, as we pointed out above, their method will always give the same
result, independent of the presence or abscence of an autocorrelation.

For those who like to touch with their hands, 
here is a concrete example from an event
generator code \cite{Moretto97} which, {\em by construction}, generates complex
fragment events according to 
\begin{equation}
p=\exp\left(-B/T\right)
\end{equation}
with
\begin{equation}
T=\sqrt{\frac{E}{a}},
\end{equation}
where $T$ is the temperature and $E$ is the excitation energy of the system.
In other words, the linear Arrhenius plot is contained in the events by
construction. 
\protect\underline{By using the independent variable $E$ instead of $E_t$}, 
we are
assured of \protect\underline{avoiding the
autocorrelation problem altogether}.

\begin{figure}
\centerline{\psfig{file=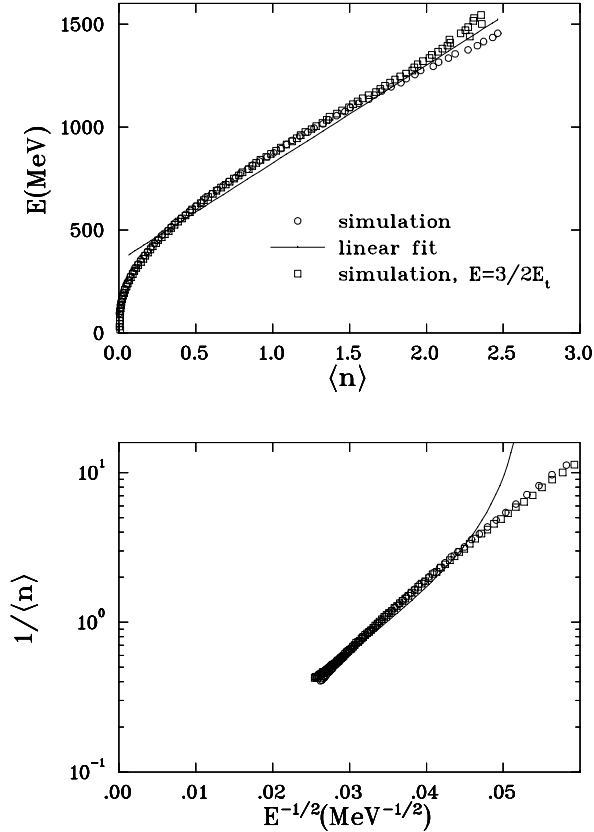,angle=90,height=11.0cm}}
\caption{Top panel: 
The excitation energy $E$ as a function of the average IMF multiplicity
from a binomial simulation 
\protect\cite{Moretto97} of the decay of a Xe
nucleus (circles). The squares show the same correlation but using for $E$ a
 scaled value
of $E_t$, $E=3/2E_t$. Bottom panel: The Arrhenius 
plot $1/\left< n\right>$ versus $1/\protect\sqrt{E}$
(circles) and $1/\left< n\right>$ versus $1/\protect\sqrt{3/2E_t}$ (squares). 
The solid line represents the (erroneous) 
``self-correlation'' explanation proposed by the authors
(see text).}
\label{fig:sim}
\end{figure}

We construct the Arrhenius plot $1/\left< n\right>$ versus $1/\sqrt{E}$ which
turns out to be linear as expected (circles in the 
bottom panel of Fig.~\ref{fig:sim}). 
Then we notice that $E$ and $\left<
n\right>$ can be plotted one against the other (circles in the top panel of
Fig.~\ref{fig:sim}) and we approximate 
\begin{equation}
E=a+b\left< n\right>
\label{eq:Estar}
\end{equation}
by fitting the dependence with a straight line. Then we substitute $E$ with the
value given by Eq.~(\ref{eq:Estar}) and note, as the authors, that the
Arrhenius plot constructed from this substitution (solid line in bottom panel
of Fig.~\ref{fig:sim}) 
remains linear over the range in 
$E_t$ where Eq.~(\ref{eq:Estar}) well represents
the simulation. Hopefully by now the game has become rather
tedious,
and nobody would argue that the result is just due to an autocorrelation.

To verify the extent of autocorrelations in this model, we follow the same
procedure as above but use $E_t$, like in the data. Now, we expect some
autocorrelation. The question is: how much? We obtain a 
linear Arrhenius plot, whose slope should coincide with the slope of the
previous plot after scaling $E_t$ by 3/2. (The model assumes $E_t=2/3E$.) 
Indeed it does so to
within better than 10 percent \cite{Moretto97} (open squares of
Fig.~\ref{fig:sim}) 
{\em This difference in slope is the extent of the autocorrelation problem}. By
redefining $\left< n\right>$ through an increase of the element threshold 
($Z_{\rm th}$) in the
definition of IMF (IMF: $Z_{\rm th}\le Z\le 20$), we can substantially 
reduce even this small
autocorrelation. Yet the authors would conclude that in both cases, using $E$
or $E_t$, the linear Arrhenius plots are completely due to autocorrelation.

\begin{figure}
\centerline{\psfig{file=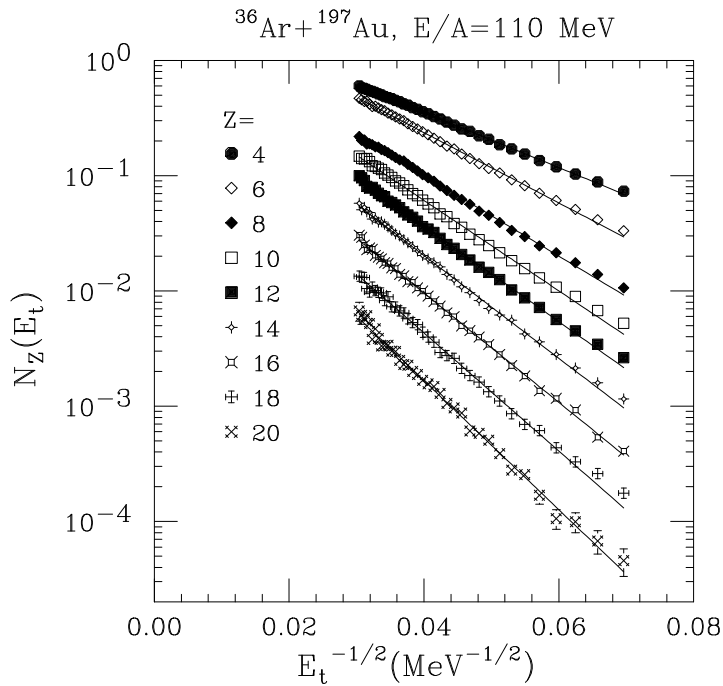,angle=90,height=9.0cm}}
\caption{The average yield per event of the different indicated elements
(symbols) as a function of $1/\protect\sqrt{E_t}$. The lines are linear fits to
the data.
}
\label{fig:ave_n_arrhenius}
\end{figure}

An even better procedure, which we can 
apply directly to experimental data, is the
following.
We can redefine the IMFs to be the fragments of a single element, say
$Z$=10. Now the contribution of the $Z$=10 fragments to $E_t$ is so small that
it can be neglected and 
\protect\underline{consequently autocorrelation is negligible} as well. 
Because of the restriction of the definition of an IMF (one single $Z$),
the probability of emitting an IMF becomes very small, and in fact,
we observe that 
the distribution of multiplicities becomes Poissonian. The Arrhenius
plots of $\log\left< n\right>$ versus $1/\sqrt{E_t}$ remain linear and the
linearity
can be tested for each $Z$ as in Fig.~\ref{fig:ave_n_arrhenius} 
without the worry of
autocorrelation. The remarkable set of straight lines shown in
Fig.~\ref{fig:ave_n_arrhenius}, together with many other sets obtained for all
the experiments we have analyzed, shows that 
thermal scaling is a {\em real physical feature} quite safe from the dreaded
fangs of the Uroboros snake.
Yet the procedure employed by the
authors will \underline{always} yield an autocorrelation, regardless of whether
one is present or not.

We now return to the first criticism, on which there is little to say.

The authors claim that there is no {\em a priori} reason for choosing a
binomial instead of a constrained Poissonian, because they both fit the
experimental multiplicity distributions equally well. We have no great argument
with this. We actually like the Poissonian case 
and we would have chosen it if the
data had not indicated a very high probability and a constraint (in size or
temporal) to boot \cite{Moretto97}.  A binomial satisfied both requirements. By
requiring a fixed number of tosses $m$, it incorporates both a scale and a
constraint. There are other ways to achieve the same goal. For instance, we
could chop the Poissonian above a prescribed $n$ or do something else, like the
authors did. In point of fact, when we restrict the definition of complex
fragments to a single atomic number $Z$ (see Fig.~\ref{fig:ave_n_arrhenius}), 
the probability falls so much that the
``binomials'' become Poissonian on their own, removing us from the
embarrassment of choice. 

{\em But} binomials, Poissonians, and constrained Poissonians, 
all share the feature
of being constructed upon {\em one} fragment probabilities. All of them
construct the $n$-fragment probability $P_n$ from the one fragment probability
$p$. All are {\em reducible} and 
this is our experimental point! A Poissonian is
{\em quintessentially reducible}. 
Because of the loss of scale, its elementary
probability is arbitrarily normalizable and coincides with the mean.
Remembering that the Poissonian originates from the binomial, we get 
$\left< n\right>=mp$ without however, being able to disentangle $m$ from $p$. On
the other hand, if we introduce a constraint, 
in whichever expression, we get back the
approximate scale $m\approx Z_0/\left< Z\right>$ where $Z_0$ is the source size.

\begin{figure}
\centerline{\psfig{file=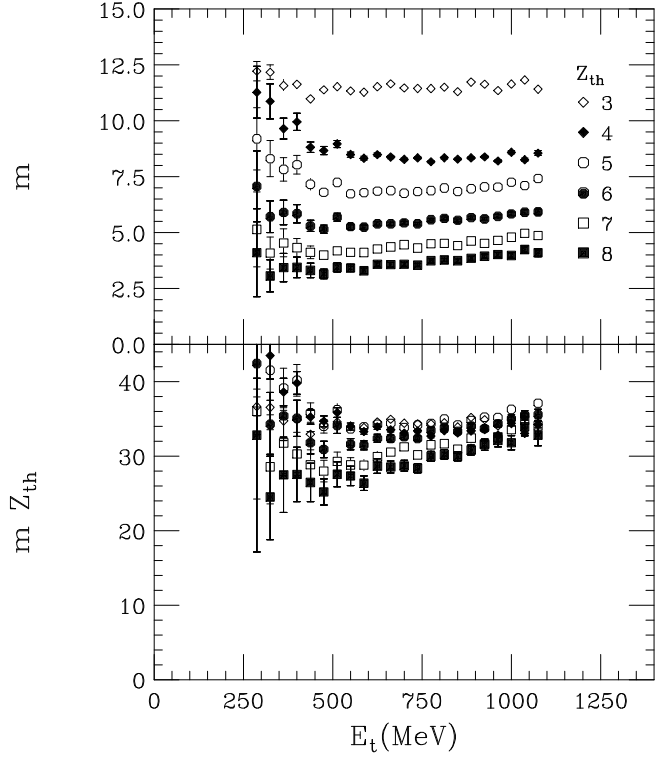,height=10.0cm,angle=90}}
\caption{The empirical scaling of $mZ_{\rm th}$ for
different $Z_{\rm th}$ (different
symbols) for the reaction $^{36}$Ar+$^{197}$Au at $E/A$=110 MeV.}
\label{fig:mscale}
\end{figure}

An example of a mass constraint in our binomial fits 
is given by the approximate
scaling that we have observed \cite{Moretto97} in the experiments:
\begin{equation}
m_{Z_{\rm th}}Z_{\rm th}\approx {\rm constant}
\end{equation}
shown in Fig.~\ref{fig:mscale}.

In view of the above, we do not see any merit to the authors criticism. 
Their function is
reducible by construction, since it assumes {\em a priori} 
independent emission (see their appendix).
Furthermore, their constraint reintroduces the scale, by the back 
door, that the Poissonian threw
out the window. Reducibility is our key statement, binomial or Poissonian, as
the case may be. 

The authors seem to imply a big difference between the two,
but the issue is reducibility.
Clearly, our experimental claim that the $n$-fragment multiplicity
distributions are \underline{reducible} to a single one fragment 
probability is not ever
called into question by the authors.

In summary, the present preprint \cite{Tsa97} 
attempts to give alternative or trivial
explanations for the reducible
and ``thermal'' nature of the experimental IMF excitation functions reported in
\cite{Moretto95,Tso95,Moretto97}. We 
have demonstrated that the proposed ``self-correlation''
explanation for linear Arrhenius plots is based upon a flawed autocorrelation
analysis involving \protect\underline{circular reasoning}. Equivalently, 
the authors have used a total derivative when they should have taken a
partial derivative. This is a fatal mistake which produces the
``self-correlation'' result, independent of whether
or not
there are autocorrelations truly present.

\end{document}